\documentclass[aps,prb,reprint,superscriptaddress,preprintnumbers]{revtex4-2}
\usepackage[T1]{fontenc}
\usepackage{times}
\usepackage{graphicx,color}
\usepackage{amsfonts,amsmath,amssymb,amsbsy}
\usepackage[colorlinks=true,linkcolor=blue,citecolor=blue,urlcolor=blue]{hyperref}

\let\mathbf=\boldsymbol

\def\blue#1{\textcolor{blue}{#1}}
\def\emph#1{\textcolor{magenta}{#1}}
\begin{document}

\title{Reversible magnetic domain reorientation induced by magnetic field pulses with fixed direction}

\author{Xichao Zhang}
\thanks{These authors contributed equally to this work.}
\affiliation{Department of Applied Physics, Waseda University, Okubo, Shinjuku-ku, Tokyo 169-8555, Japan}

\author{Jing Xia}
\thanks{These authors contributed equally to this work.}
\affiliation{Department of Electrical and Computer Engineering, Shinshu University, 4-17-1 Wakasato, Nagano 380-8553, Japan}

\author{Oleg A. Tretiakov}
\affiliation{School of Physics, The University of New South Wales, Sydney 2052, Australia}

\author{Guoping Zhao}
\affiliation{College of Physics and Electronic Engineering, Sichuan Normal University, Chengdu 610068, China}

\author{Yan Zhou}
\affiliation{School of Science and Engineering, The Chinese University of Hong Kong, Shenzhen, Guangdong 518172, China}

\author{\\ Masahito Mochizuki}
\email[Email:~]{masa_mochizuki@waseda.jp}
\affiliation{Department of Applied Physics, Waseda University, Okubo, Shinjuku-ku, Tokyo 169-8555, Japan}

\author{Xiaoxi Liu}
\email[Email:~]{liu@cs.shinshu-u.ac.jp}
\affiliation{Department of Electrical and Computer Engineering, Shinshu University, 4-17-1 Wakasato, Nagano 380-8553, Japan}

\author{Motohiko Ezawa}
\email[Email:~]{ezawa@ap.t.u-tokyo.ac.jp}
\affiliation{Department of Applied Physics, The University of Tokyo, 7-3-1 Hongo, Tokyo 113-8656, Japan}

\begin{abstract}
Nanoscale magnetic domains with controllable configurations could be used for classical and quantum applications, where the switching of magnetization configurations is an essential operation for information processing. Here, we report that the magnetic domain reorientation in a notched ferromagnetic nanotrack can be realized and effectively controlled by applying uniform magnetic field pulses in a fixed in-plane direction perpendicular to the nanotrack. Our micromagnetic simulation results show that the configurations of magnetic domains in the notched nanotrack can be switched between a head-to-head state and a tail-to-tail state in a reversible manner driven by magnetic field pulses, while it is unnecessary to reverse the direction of the magnetic field. Such a unique magnetic domain reorientation dynamics is found to depend on magnetic parameters and nanotrack geometries. The reorientation dynamics of magnetic domains also depends on the strength and length of the applied magnetic field pulse. In addition, we point out that the notches at the center of the nanotrack play an important role for the stabilization of the head-to-head and tail-to-tail states during the magnetic domain reorientation. We also qualitatively explain the field-induced reorientation phenomenon with a simplified two-dimensional macrospin model. Our results may make it possible to build spintronic devices driven by a fixed magnetic field. Our findings may also motivate future studies to investigate the classical and quantum applications based on nanoscale magnetic domains.
\end{abstract}

\date{July 28, 2023}


\maketitle

\section{Introduction}
\label{se:Introduction}

The magnetization configurations in nanoscale ferromagnetic tracks can be used to store binary information and performing storage functions~\cite{Chappert_2007,Brataas_2012,Parkin_NNANO2015A,Kumar_IEEETMAG2019,Parkin_PIEEE2020}.
For example, the magnetic domain walls in a nanotrack with either in-plane or out-of-plane magnetization can be driven into motion by magnetic and electric means~\cite{Schryer_1974,Yamaguchi_2004,Tatara_2004,Thiaville_2005,Ralph_2008,Parkin_SCIENCE2008,Hayashi_2008,Thiaville_2012,Emori_2013,Parkin_NNANO2015B,Kumar_2021,Tretiakov_PRL2008,Tretiakov_PRB2008,Tretiakov_PRL2010,Tretiakov_PRL2012,Tretiakov_APL2016,Yamanouchi_Nature2004}, which is essential for the racetrack-type memory and relevant logic computing devices~\cite{Chappert_2007,Brataas_2012,Parkin_NNANO2015A,Kumar_IEEETMAG2019,Parkin_PIEEE2020}.
Similarly, chiral skyrmionic textures stabilized in nanotracks may also be used for information storage and processing~\cite{Nagaosa_NNANO2013,Mochizuki_Review,Finocchio_JPD2016,Kang_PIEEE2016,Fert_NATREVMAT2017,Everschor_JAP2018,Zhang_JPCM2020,Gobel_PP2021,Luo_APLM2021,Marrows_APL2021,Reichhardt_2021,Wiesendanger_NATREVMAT2016,Wanjun_PHYSREP2017,Back_JPD2020}.
The racetrack-type memory applications based on magnetic domain walls could be driven by the magnetic field and electric current~\cite{Schryer_1974,Yamaguchi_2004,Tatara_2004,Thiaville_2005,Ralph_2008,Parkin_SCIENCE2008,Hayashi_2008,Thiaville_2012,Emori_2013,Parkin_NNANO2015B,Kumar_2021,Chappert_2007,Brataas_2012,Parkin_NNANO2015A,Kumar_IEEETMAG2019,Parkin_PIEEE2020,Tretiakov_PRL2008,Tretiakov_PRB2008,Tretiakov_PRL2010,Tretiakov_PRL2012,Tretiakov_APL2016,Yamanouchi_Nature2004}, while the skyrmion-based one is usually driven by the electric current as the skyrmion cannot be driven into translational motion by a uniform and constant magnetic field~\cite{Nagaosa_NNANO2013,Mochizuki_Review,Finocchio_JPD2016,Kang_PIEEE2016,Fert_NATREVMAT2017,Everschor_JAP2018,Zhang_JPCM2020,Gobel_PP2021,Luo_APLM2021,Marrows_APL2021,Reichhardt_2021,Wiesendanger_NATREVMAT2016,Wanjun_PHYSREP2017,Back_JPD2020}.

In order to realize next-generation spintronic applications based on nonvolatile magnetization configurations, it is still important to explore and understand the dynamics of magnetization configurations induced by the most fundamental external stimuli, such as the magnetic field and electric current.
The dynamics of ferromagnetic domain walls in nanotracks controlled by magnetic field pulses has been investigated for decades, however, most published reports are focused on the field-induced or current-induced motion of domain walls~\cite{Schryer_1974,Yamaguchi_2004,Tatara_2004,Thiaville_2005,Ralph_2008,Parkin_SCIENCE2008,Hayashi_2008,Thiaville_2012,Emori_2013,Parkin_NNANO2015B,Kumar_2021,Chappert_2007,Brataas_2012,Parkin_NNANO2015A,Kumar_IEEETMAG2019,Parkin_PIEEE2020,Tretiakov_PRL2008,Tretiakov_PRB2008,Tretiakov_PRL2010,Tretiakov_PRL2012,Tretiakov_APL2016,Yamanouchi_Nature2004}.
In a racetrack-type memory device, the field-induced in-line motion of domain walls is a key operation for manipulating the encoded data on the nanotrack, however, domain walls will be erased and re-created during the write-in and read-out operations~\cite{Parkin_NNANO2015A,Kumar_IEEETMAG2019,Parkin_PIEEE2020}.
In the same manner, the chiral skyrmionic textures used in racetrack-type applications may also need to be frequently created and deleted during the write-in and read-out operations~\cite{Finocchio_JPD2016,Kang_PIEEE2016,Fert_NATREVMAT2017,Everschor_JAP2018,Zhang_JPCM2020,Gobel_PP2021,Luo_APLM2021,Marrows_APL2021}.

A possible solution to avoid frequent creation and deletion of domain walls in magnetic memory and logic computing applications is to utilize the magnetic domain reorientation and switching dynamics in well-designed nanostructures~\cite{Allwood_2005,Ikeda_2010,Miron_2011,Liu_2012,Bhatti_2017,Albrecht_2015}.
Namely, the data write-in and read-out process is based on the manipulation of the domain configuration instead of the domain wall position~\cite{Allwood_2005,Ikeda_2010,Miron_2011,Liu_2012,Bhatti_2017,Albrecht_2015}, because that each data bit is stored in a single isolated nanostructure or region so that it will be unnecessary to frequently create, move, and delete domain walls as the domain wall position can be fixed during the magnetic domain reorientation.
However, the magnetic domain reorientation dynamics driven by magnetic field pulses in ferromagnetic nanotracks is still elusive, however, the magnetic domain reorientation dynamics could be vital for the design and development of future applications.

In this paper, we report the magnetic domain reorientation dynamics in a notched ferromagnetic nanotrack driven by uniform in-plane magnetic field pulses perpendicular to the nanotrack, where the configurations of magnetic domains could be reversibly switched by magnetic field pulses without reversing the field direction.
Our results could be useful for the design and development of novel information processing applications based on the manipulation of magnetic domains in nanoscale tracks with modified geometries.

\section{Methods}
\label{se:Methods}

In this work, we consider an ultra-thin ferromagnetic nanotrack with two square notches at the center of the upper and lower edges along the length direction, as shown in Fig.~\ref{FIG4}(a).
The default length of the nanotrack in the $x$ direction is equal to $100$ nm, and the default width in the $y$ direction is set to $10$ nm. The thickness of the nanotrack is fixed at $1$ nm.
The simulation is performed by using the functional micromagnetics package Object Oriented MicroMagnetic Framework (OOMMF)~\cite{OOMMF}.
The mesh size in all simulations is set to $1$ $\times$ $1$ $\times$ $1$ nm$^3$, which ensures good computational accuracy.
The open boundary conditions are applied in the $x$ and $y$ directions.

We assumed that the nanotrack is made of typical permalloy, and the magnetic parameters are~\cite{Im_2012}:
the saturation magnetization $M_\text{S}=860$ kA m$^{-1}$,
the exchange constant $A=13$ pJ m$^{-1}$,
and the magnetic anisotropy equals zero.
In the simulation, the magnetization dynamics in the permalloy nanotrack is controlled by the Landau-Lifshitz-Gilbert (LLG) equation~\cite{OOMMF},
\begin{equation}
\label{eq:LLGS-CPP}
\partial_{t}\boldsymbol{m}=-\gamma_{0}\boldsymbol{m}\times\boldsymbol{h}_{\text{eff}}+\alpha(\boldsymbol{m}\times\partial_{t}\boldsymbol{m}),
\end{equation}
where $\boldsymbol{m}$ is the reduced magnetization (i.e., $\boldsymbol{m}=\boldsymbol{M}/M_\text{S}$),
$t$ is the time,
$\gamma_0$ is the absolute gyromagnetic ratio, and 
$\alpha$ is the Gilbert damping parameter with the default value being $0.3$.
$\boldsymbol{h}_{\rm{eff}}=-\frac{1}{\mu_{0}M_{\text{S}}}\cdot\frac{\delta\varepsilon}{\delta\boldsymbol{m}}$ is the effective field, where $\mu_{0}$ and $\varepsilon$ are the vacuum permeability constant and average energy density, respectively.
The energy terms considered in the simulation include the ferromagnetic exchange energy, the applied magnetic field energy, and the demagnetization energy~\cite{OOMMF}, as expressed in the average energy density below 
\begin{equation}
\label{eq:Average-Energy} 
\varepsilon=A\left(\nabla\boldsymbol{m}\right)^{2}-\mu_{0}M_{\text{S}}\left(\boldsymbol{m}\cdot\boldsymbol{H}_{\text{a}}\right)-\frac{\mu_{0}M_{\text{S}}}{2}\left(\boldsymbol{m}\cdot\boldsymbol{H}_{\text{d}}\right),
\end{equation}
where $A$ is the ferromagnetic exchange constant, $\boldsymbol{H}_{\text{a}}$ is the applied magnetic field, and $\boldsymbol{H}_{\text{d}}$ is the demagnetization field.
Note that the magnetic anisotropy energy equals zero in our model as the permalloy nanotrack has no crystalline magnetic anisotropy~\cite{Im_2012}, while the demagnetization effect may result in certain magnetic shape anisotropy.

\begin{figure}[t]
\centerline{\includegraphics[width=0.43\textwidth]{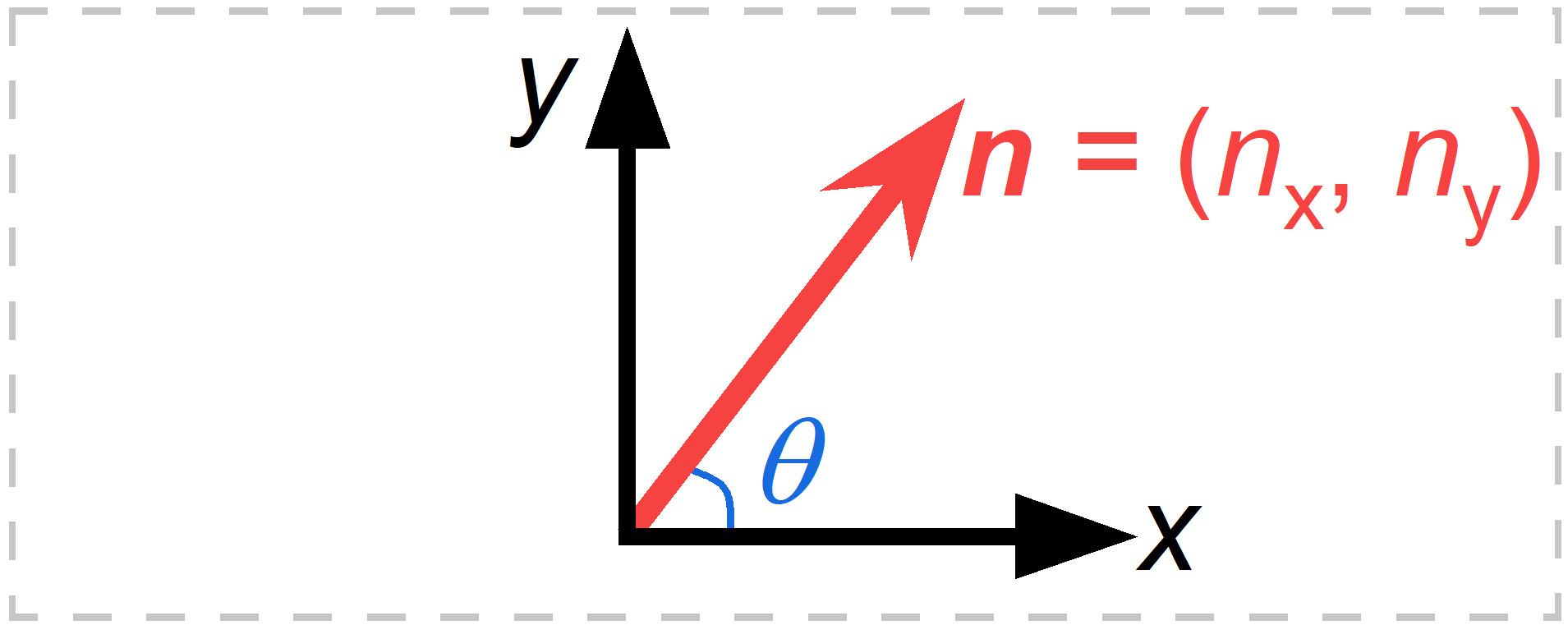}}
\caption{%
Schematic illustration of a two-dimensional macrospin model describing the a single spin driven by a magnetic field pulse. The angle between the spin vector direction and the $+x$ direction is defined as $\theta$. The magnetic field pulse is applied in the $+y$ direction.
}
\label{FIG1}
\end{figure}

\begin{figure}[t]
\centerline{\includegraphics[width=0.49\textwidth]{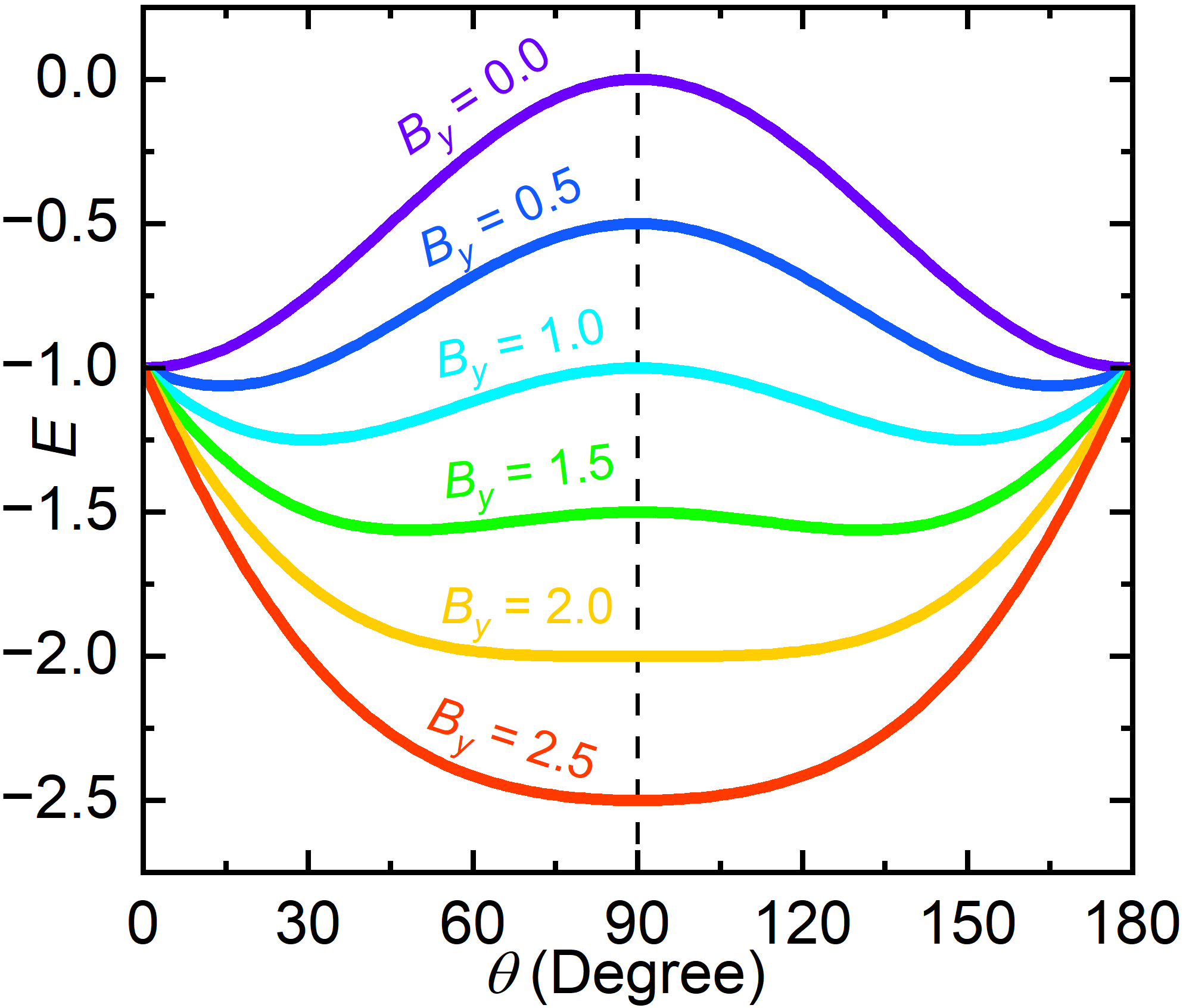}}
\caption{%
The total energy of the macrospin system as a function of $\theta$ for $K=1$ and $B_{y}=0-2.5$. $K$ and $B_{y}$ are simplified parameters, which control the shape anisotropy magnitude and magnetic field strength, respectively.
}
\label{FIG2}
\end{figure}

\begin{figure}[t]
\centerline{\includegraphics[width=0.49\textwidth]{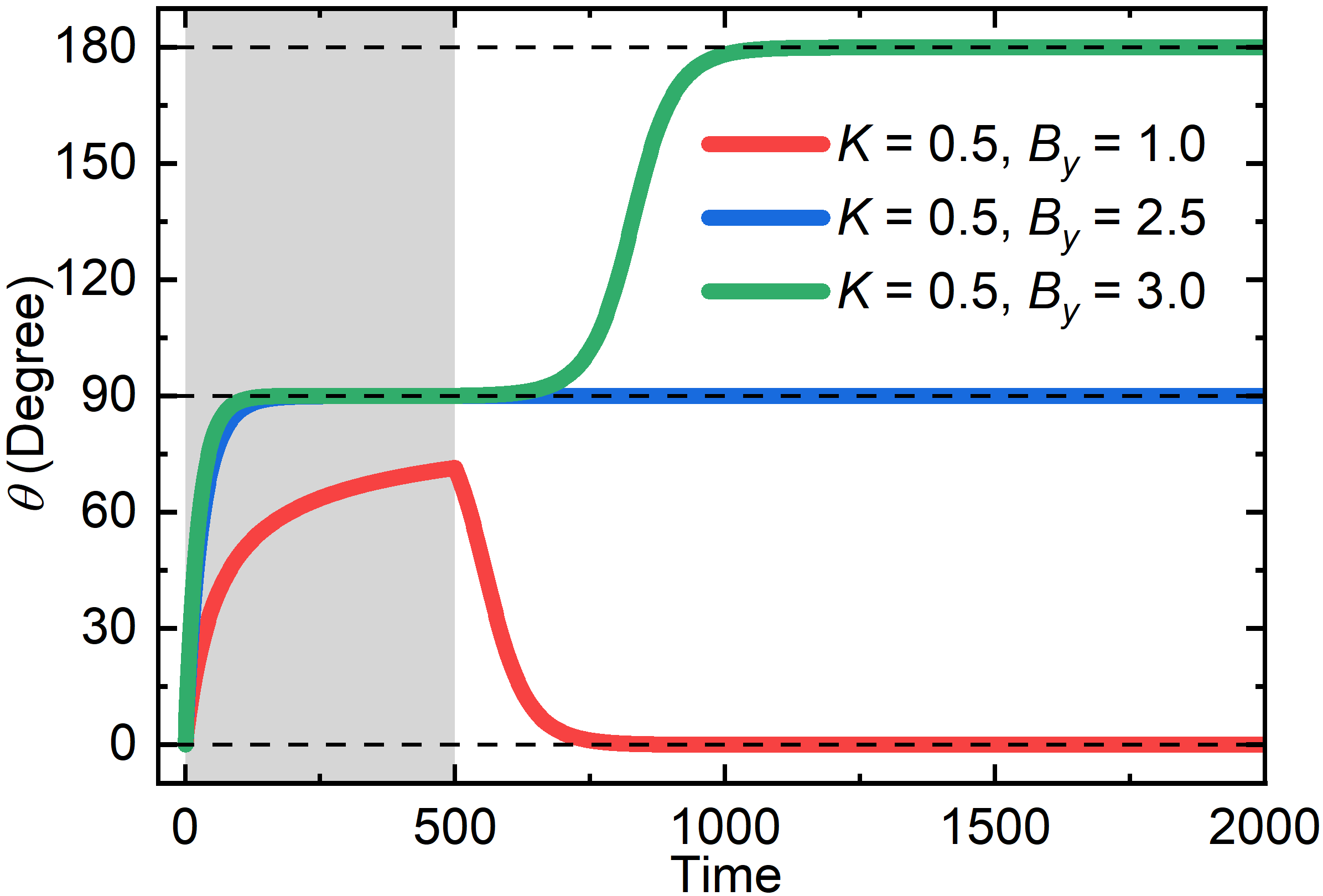}}
\caption{%
Field-induced reorientation of a single macrospin. $\theta$ as functions of time for $B_{y}=1.0$, $2.5$, and $3.0$. The shape magnetic anisotropy strength is fixed at $K=0.5$. The magnetic field is applied during $t=0-500$, as indicated by the gray background.
}
\label{FIG3}
\end{figure}

\begin{figure*}[t]
\centerline{\includegraphics[width=0.83\textwidth]{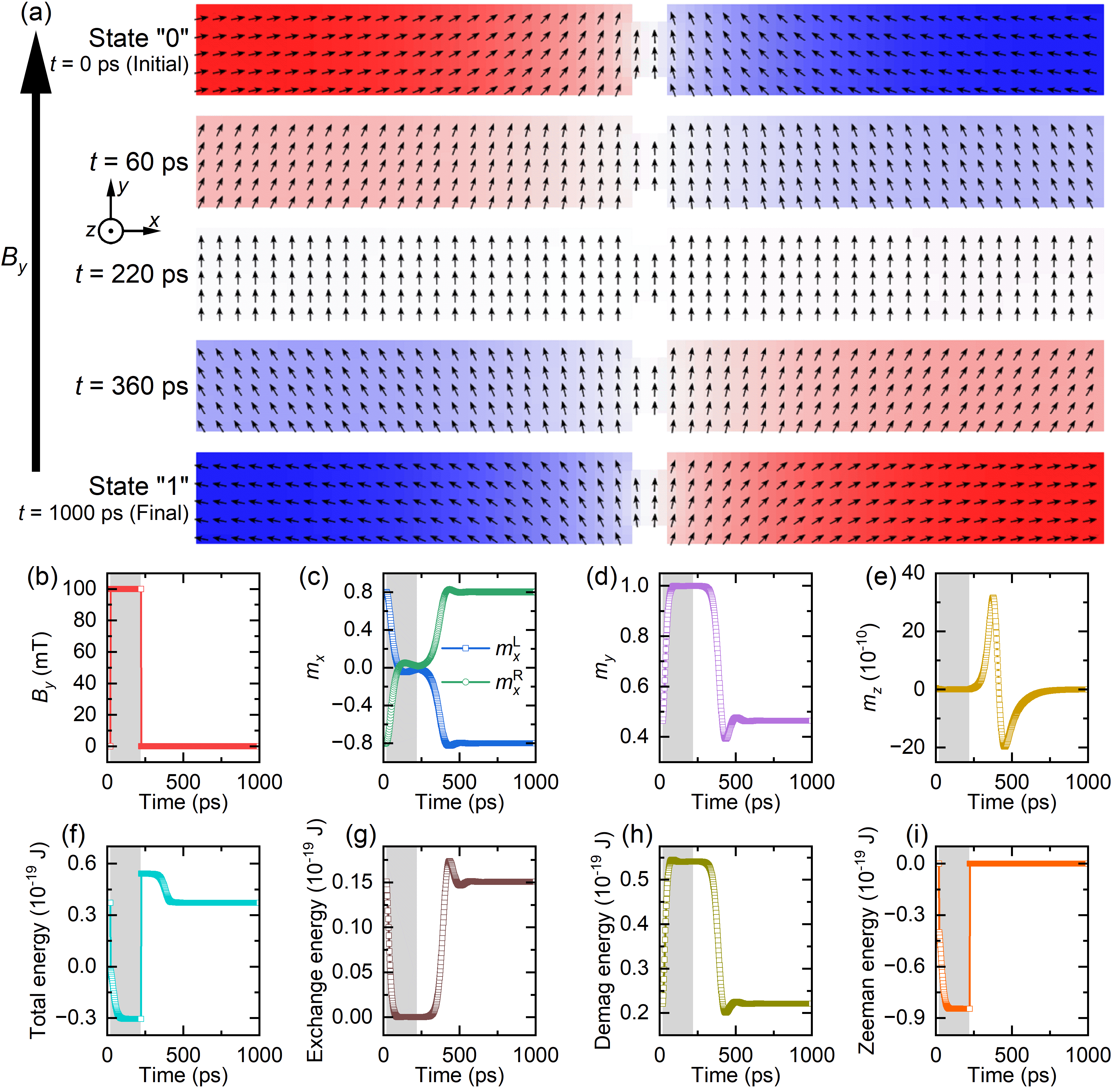}}
\caption{%
Field-induced reorientation of magnetic domains in a notched nanotrack leading to the switching from state ``0'' to state ``1''.
(a) Top-view snapshots showing the magnetization configurations in the notched nanotrack at selected times. The arrows represent the magnetization directions. The in-plane magnetization component ($m_x$) is color coded: red is pointing at the $+x$ direction, blue is pointing at the $-x$ direction, and white is pointing at the $\pm y$ directions. The initial state ``0'' is a metastable head-to-head magnetization configuration, while the final state ``1'' is a metastable tail-to-tail magnetization configuration. The switching from state ``0'' to state ``1'' is realized by the magnetization reorientation driven by a magnetic field pulse applied at the $+y$ direction. The field strength $B_y=100$ mT, and the pulse length $\tau=200$ ps (i.e., applied during $t=20-220$ ps). The system is relaxed until $t=1000$ ps. Here, $\alpha=0.3$, $A=13$ pJ m$^{-1}$, and $M_\text{S}=860$ kA m$^{-1}$.
(b) The applied magnetic field strength $B_y$ as a function of time.
(c) The reduced in-plane magnetization components for the left half ($m_{x}^{\text{L}}$) and right half ($m_{x}^{\text{R}}$) of the nanotrack as functions of time.
(d) The reduced in-plane magnetization component $m_y$ as a function of time.
(e) The reduced out-of-plane magnetization component $m_z$ as a function of time.
(f) The total energy as a function of time.
(g) The exchange energy as a function of time.
(h) The demagnetization energy as a function of time.
(i) The applied magnetic field energy (i.e., Zeeman energy) as a function of time.
The magnetic field pulse duration is indicated by the gray background.
}
\label{FIG4}
\end{figure*}

\begin{figure*}[t]
\centerline{\includegraphics[width=0.83\textwidth]{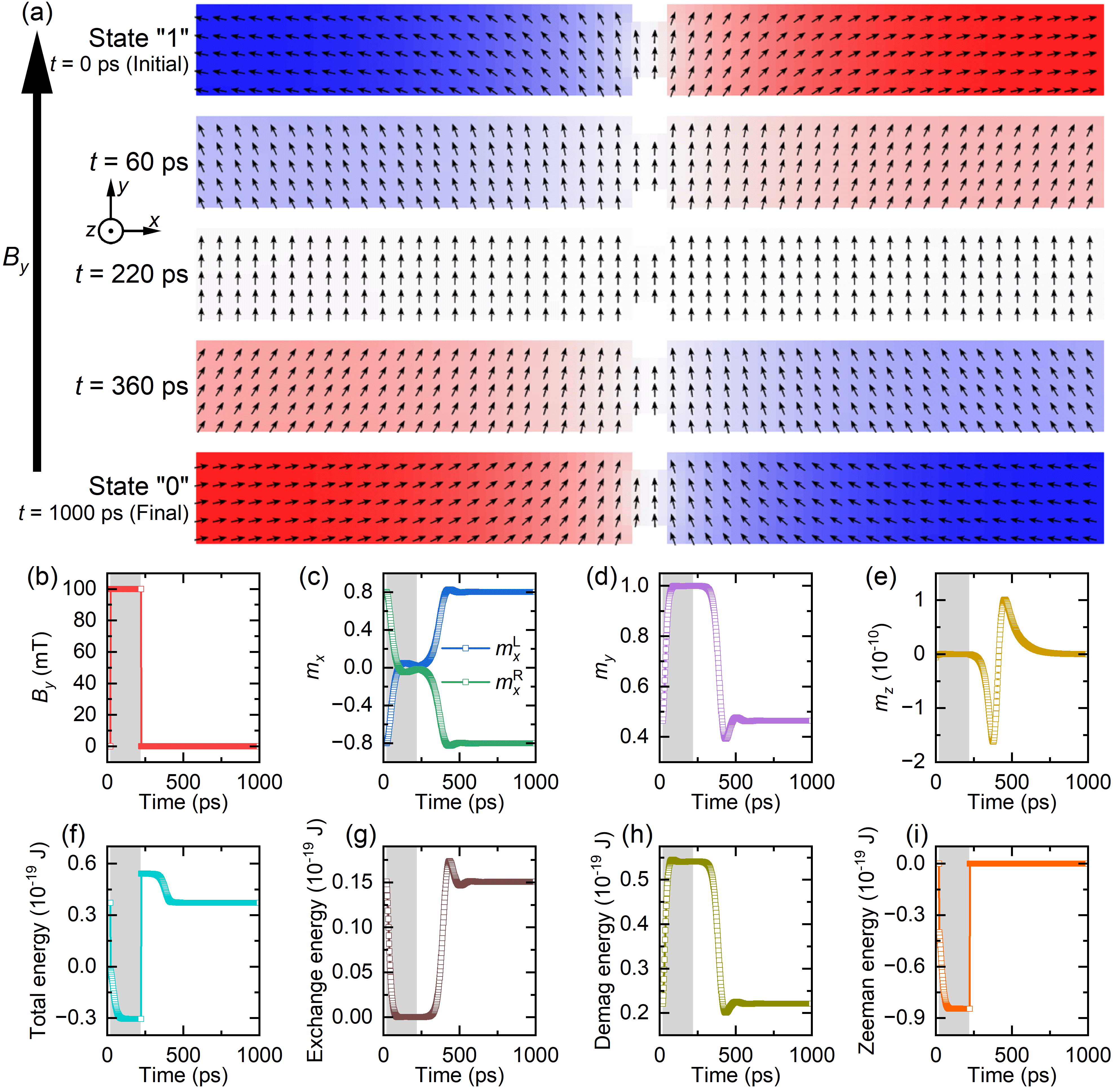}}
\caption{%
Field-induced reorientation of magnetic domains in a notched nanotrack leading to the switching from state ``1'' to state ``0''.
(a) Top-view snapshots showing the magnetization configurations in the notched nanotrack at selected times. The arrows represent the magnetization directions. The in-plane magnetization component ($m_x$) is color coded: red is pointing at the $+x$ direction, blue is pointing at the $-x$ direction, and white is pointing at the $\pm y$ directions. The initial state ``1'' is a metastable tail-to-tail magnetization configuration, while the final state ``0'' is a metastable head-to-head magnetization configuration. The switching from state ``1'' to state ``0'' is realized by the magnetization reorientation driven by a magnetic field pulse applied at the $+y$ direction. The field strength $B_y=100$ mT, and the pulse length $\tau=200$ ps (i.e., applied during $t=20-220$ ps). The system is relaxed until $t=1000$ ps. Here, $\alpha=0.3$, $A=13$ pJ m$^{-1}$, and $M_\text{S}=860$ kA m$^{-1}$.
(b) The applied magnetic field strength $B_y$ as a function of time.
(c) The reduced in-plane magnetization components for the left half ($m_{x}^{\text{L}}$) and right half ($m_{x}^{\text{R}}$) of the nanotrack as functions of time.
(d) The reduced in-plane magnetization component $m_y$ as a function of time.
(e) The reduced out-of-plane magnetization component $m_z$ as a function of time.
(f) The total energy as a function of time.
(g) The exchange energy as a function of time.
(h) The demagnetization energy as a function of time.
(i) The applied magnetic field energy (i.e., Zeeman energy) as a function of time.
The magnetic field pulse duration is indicated by the gray background.
}
\label{FIG5}
\end{figure*}

\begin{figure}[t]
\centerline{\includegraphics[width=0.50\textwidth]{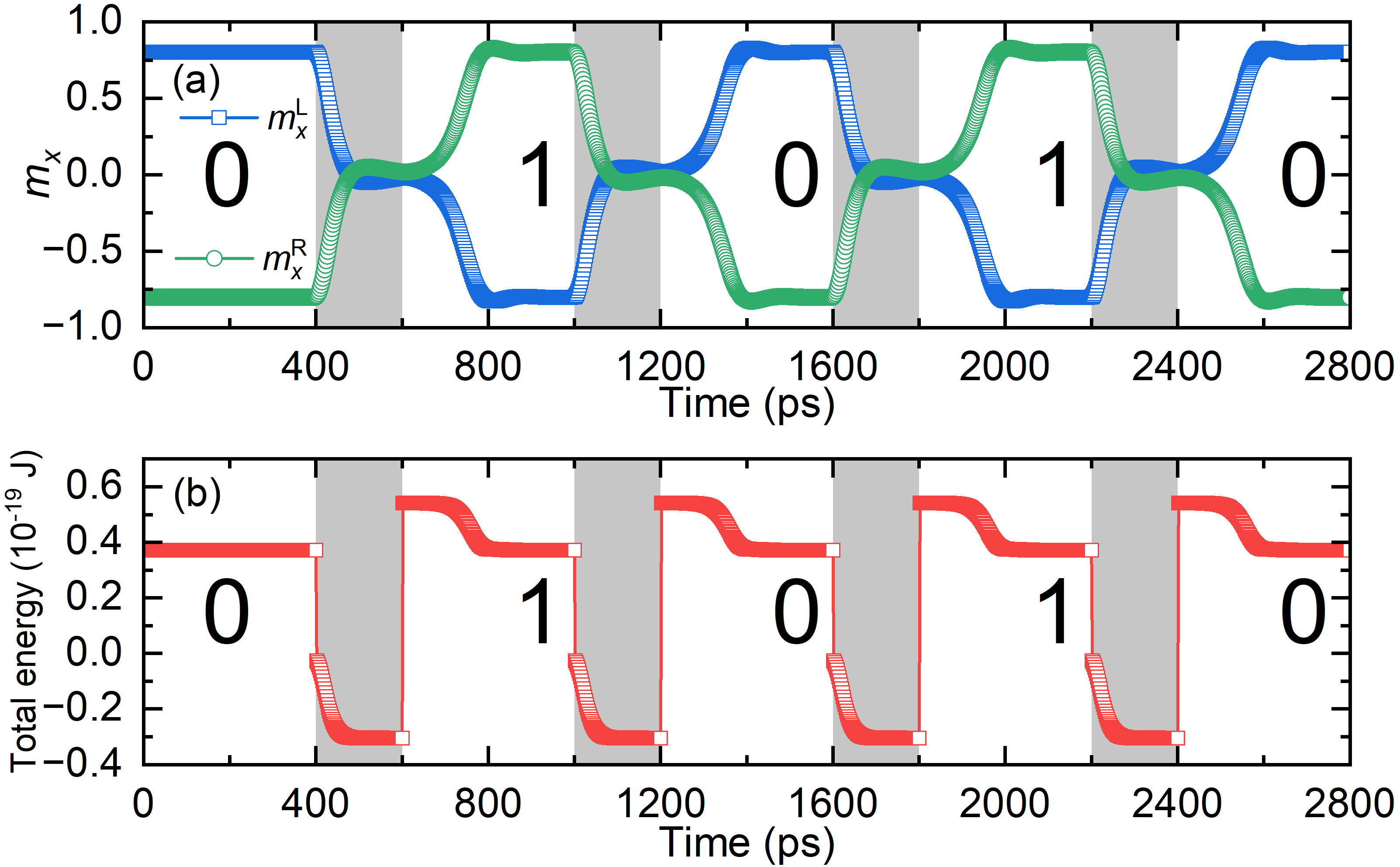}}
\caption{%
Reversible reorientation of magnetic domains in a notched nanotrack leading to the switching between state ``0'' and state ``1'' in a controlled manner.
(a) The reduced in-plane magnetization components for the left half ($m_{x}^{\text{L}}$) and right half ($m_{x}^{\text{R}}$) of the nanotrack as functions of time.
(b) The total energy as a function of time.
Here, $\alpha=0.3$, $A=13$ pJ m$^{-1}$, and $M_\text{S}=860$ kA m$^{-1}$. The field strength $B_y=100$ mT. The pulse length $\tau=200$ ps, which is indicated by the gray background.
}
\label{FIG6}
\end{figure}

\begin{figure}[t]
\centerline{\includegraphics[width=0.49\textwidth]{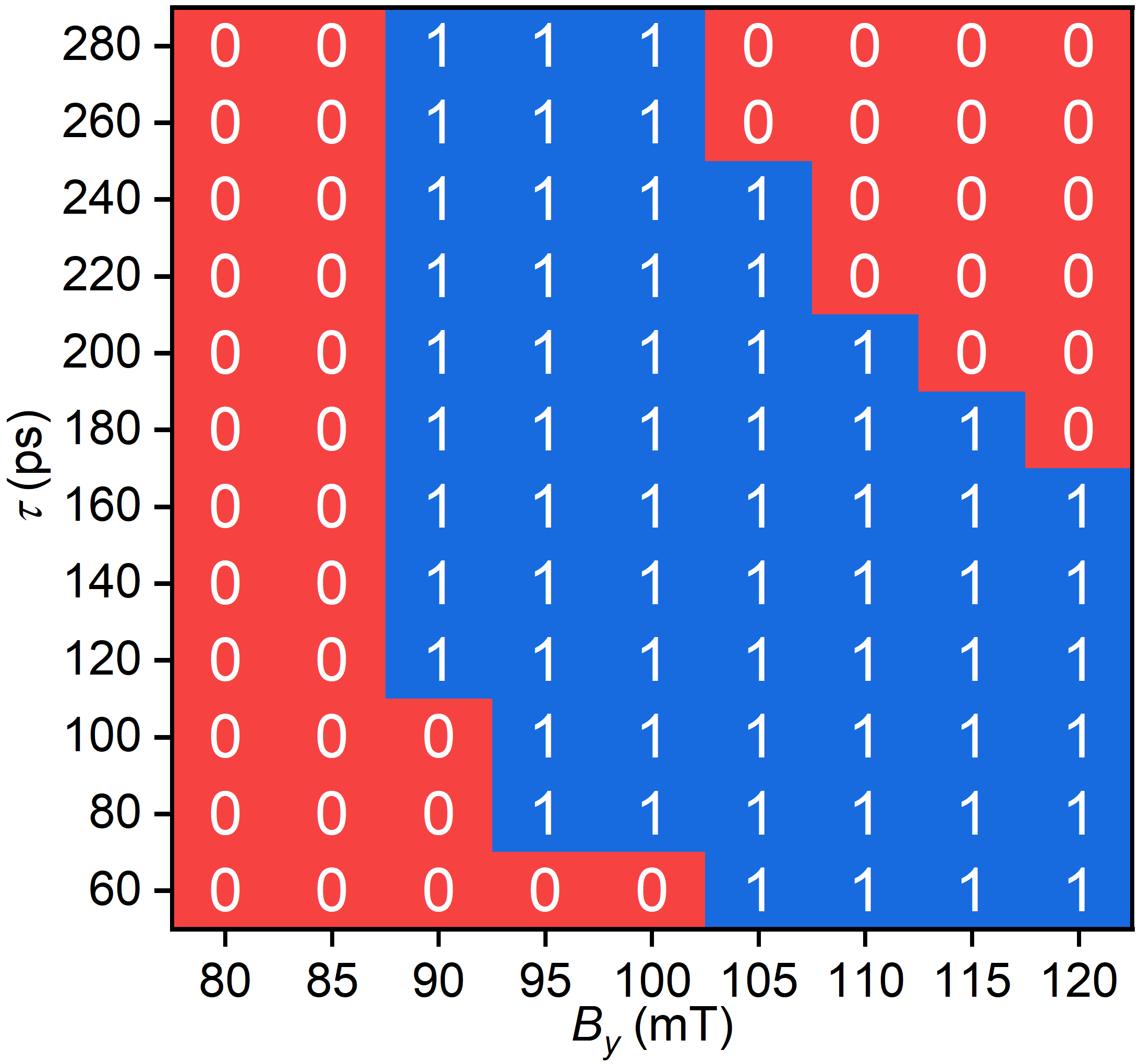}}
\caption{%
Switching phase diagram of a notched nanotrack driven by a single magnetic field pulse for different pulse length $\tau$ and field strength $B_y$.
The initial relaxed state before the pulse application is the state ``0'' with a head-to-head magnetization configuration. The magnetic field pulse is applied at the $+y$ direction. The switching from state ``0'' to state ``1'' is realized for certain values of $\tau$ and $B_y$. Here, $\alpha=0.3$, $A=13$ pJ m$^{-1}$, and $M_\text{S}=860$ kA m$^{-1}$.
}
\label{FIG7}
\end{figure}

\begin{figure}[t]
\centerline{\includegraphics[width=0.49\textwidth]{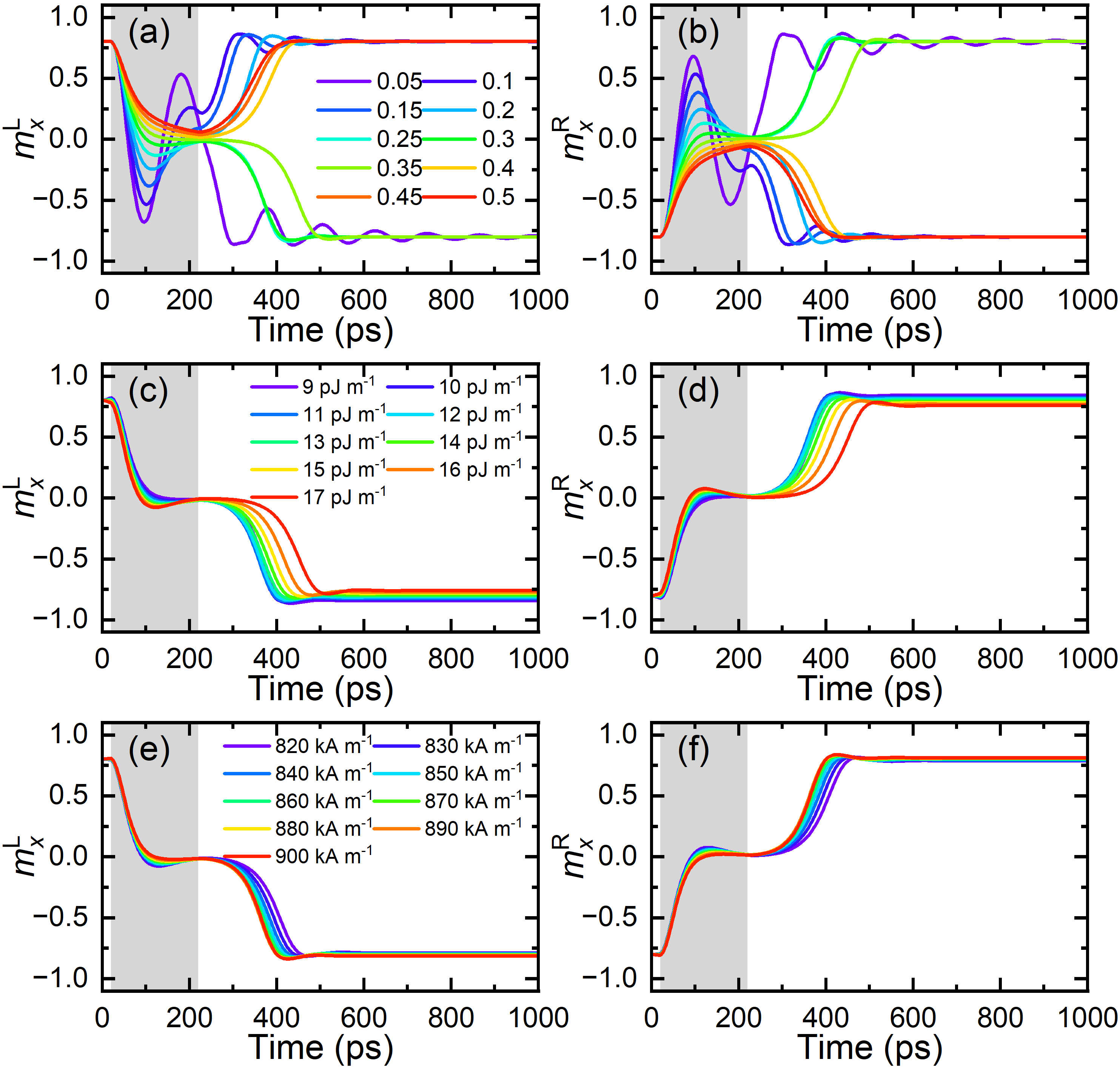}}
\caption{%
Dynamics of magnetic domains driven by a single magnetic field pulse in notched nanotracks with different material parameters.
(a) Time-dependent reduced in-plane magnetization component for the left half of the nanotrack $m_{x}^{\text{L}}$ for different damping parameter $\alpha$.
(b) Time-dependent reduced in-plane magnetization component for the right half of the nanotrack $m_{x}^{\text{R}}$ for different $\alpha$.
(c) Time-dependent $m_{x}^{\text{L}}$ for different exchange constant $A$.
(d) Time-dependent $m_{x}^{\text{R}}$ for different $A$.
(e) Time-dependent $m_{x}^{\text{L}}$ for different saturation magnetization $M_\text{S}$.
(f) Time-dependent $m_{x}^{\text{R}}$ for different $M_\text{S}$.
Here, the initial relaxed state before the pulse application is the state ``0'' with a head-to-head magnetization configuration. The magnetic field pulse is applied at the $+y$ direction. The field strength $B_y=100$ mT, and the pulse length $\tau=200$ ps (i.e., applied during $t=20-220$ ps as indicated by the gray background). The system is relaxed until $t=1000$ ps.
}
\label{FIG8}
\end{figure}

\section{Results and Discussion}
\label{se:Results}

\subsection{A two-dimensional macrospin model}
\label{se:Macrospin}

Before discussing the micromagnetic simulation results, we first analyze the magnetic field-induced switching of the in-plane magnetization in a nanotrack with a simplified two-dimensional macrospin model.
As depicted in Fig.~\ref{FIG1}, we neglect the spatial variation of the magnetization within the left half of the nanotrack and assume that the magnetization is described by a two-dimensional single macroscopic spin vector $\boldsymbol{n}=(n_x, n_y)$. We define the angle between the spin vector direction and the $+x$ direction as $\theta$, and therefore, the spin vector could be expressed as
\begin{equation}
\label{eq:Spin}
\boldsymbol{n}=\hat{x}\cos\theta+\hat{y}\sin\theta.
\end{equation}
To mimic the demagnetization effect, we consider a uniaxial shape magnetic anisotropy with the $x$ axis being the easy axis. We also consider a magnetic field pulse applied in the $+y$ direction. Thus, the energy of the system is given as
\begin{equation}
\label{eq:Macrospin-Energy}
E=-K\cos^{2}\theta-B_{y}\sin\theta,
\end{equation}
where $K$ and $B_{y}$ are simplified parameters that control the anisotropy magnitude and magnetic field strength, respectively.
In Fig.~\ref{FIG2}, it can be seen that the total energy $E$ reaches its maximum value when $\theta=90^{\circ}$ in the absence of the magnetic field (i.e., $B_{y}=0$), while it reaches its minimum value when $\theta=90^{\circ}$ upon the application of a strong magnetic field (e.g., $B_{y}=2.5$). However, it should be noted that the energy minimum may be found at $\theta=0^{\circ}-90^{\circ}$ and $\theta=90^{\circ}-180^{\circ}$, which depends on the competition between the shape magnetic anisotropy $K$ and applied magnetic field $B_{y}$. In order to drive the switching of the spin, the strength of the applied magnetic field pulse should be large enough to create an energy profile such as the curve for $K=1$ and $B_{y}=2.5$ in Fig.~\ref{FIG2}, where $\theta$ must reach $90^{\circ}$ when the magnetic field is applied for certain time.

To qualitatively demonstrate the field-induced spin reorientation, we further obtain the Euler-Lagrange equation~\cite{EL1,EL2} for the two-dimensional macroscopic model from the Lagrangian
\begin{equation}
\label{eq:Lagrangian}
L=-\dot{\theta}+K\cos^{2}\theta+B_{y}\sin\theta,
\end{equation}
and the Euler-Lagrange equation is given as
\begin{equation}
\label{eq:Euler-Lagrange}
\frac{d}{dt}\frac{\partial L}{\partial\dot{\theta}}-\frac{\partial L}{\partial\theta}=-\frac{\partial}{\partial\dot{\theta}}\frac{\alpha}{2}\dot{\theta}^{2},
\end{equation}
with $\alpha$ being the damping parameter.
Thus, the LLG equation is obtained by polar coordinate $\theta$ as
\begin{equation}
\label{eq:LLG-Macrospin}
\dot{\theta}=-\frac{2K}{\alpha}\sin\theta\cos\theta+\frac{B_{y}}{\alpha}\cos\theta.
\end{equation}
In Fig.~\ref{FIG3}, we show three representative situations of the field-induced spin dynamics by numerically solving Eq.~(\ref{eq:LLG-Macrospin}) with the simplified assumption that $K=\alpha=0.5$.
As shown in Fig.~\ref{FIG3}, when a relatively weak magnetic field pulse of $B_{y}=1$ is applied during $t=0-500$ and the system is relaxed until $t=2000$, it can be seen that the spin vector first rotates toward the $+y$ direction (i.e., $\theta\rightarrow 90^{\circ}$) but finally relaxes to its initial direction (i.e., $\theta=0^{\circ}$). This situation suggests that a weak magnetic field pulse may not be able to trigger a full $180^{\circ}$ reorientation of the magnetization in the nanotrack.
When a stronger magnetic field pulse of $B_{y}=2.5$ is applied during $t=0-500$, it shows that the spin vector is magnetized to the $+y$ direction soon upon the application of the pulse (i.e., $\theta=90^{\circ}$). However, we note that $\theta$ remains at $90^{\circ}$ when the magnetic field pulse is turned off. The reason is due to the symmetry breaking problem, where the spin vector direction gets stuck on an unstable equilibrium. Namely, the spin vector is balanced at the maximum energy point and does not know which direction to rotate.
In micromagnetic simulations discussed below, the symmetry would be broken due to the spin precession.
Therefore, we apply a relatively strong magnetic field pulse of $B_{y}=3$ during $t=0-500$, and meanwhile, consider a tiny perturbation of the magnetic field direction to avoid the symmetry breaking problem in the macrospin model. Hence, it can be seen from Fig.~\ref{FIG3} that the spin vector first rotates to the $+y$ direction (i.e., $\theta=90^{\circ}$) when the magnetic field pulse is turned on, and then rotates to the $-x$ direction when the magnetic field pulse is turned off. This situation suggests that a strong magnetic field pulse could be able to result in a full $180^{\circ}$ reorientation of the magnetization in the nanotrack. The prerequisite is that the magnetic field pulse is strong and long enough to drive the spin to overcome the energy barrier due to the shape magnetic anisotropy.

\subsection{Reversible reorientation phenomenon}
\label{se:Reorientation}

We computationally demonstrate the typical reversible reorientation of magnetic domains in the notched nanotrack with default geometry and material parameters, where the magnetization dynamics is driven by in-plane magnetic field pulses applied in the width direction of the nanotrack, i.e., pointing at the $+y$ direction.
As shown in Fig.~\ref{FIG4}(a), the initial state in the nanotrack at $t=0$ ps is a relaxed metastable head-to-head magnetization configuration, which is defined as the state ``0'' as it could be used to carry the binary information bit ``0''. Namely, the magnetization in the left half and right half of the nanotrack are almost aligned along the $+x$ (red) and $-x$ (blue) directions, respectively.
We point out that the read-out of state ``0'' could be realized by placing two magnetic tunnel junction (MTJ) reader sensors upon the left half and right half of the nanotrack.
The two square notches at the center of the nanotrack are employed to stabilize and fix the position of a head-to-head or tail-to-tail domain wall at the nanotrack center. In this work, the default length and width of the notch are equal to $4$ nm and $2$ nm, respectively.
The notches fabricated at nanotrack edges are usually used to pin domain walls and affect the domain wall dynamics~\cite{Hayashi_2006,Atkinson_2008,Im_2009,Akerman_2010,Yuan_2014,Araujo_2019}.
The domain wall can exist in the nanotrack without the notches, however, it may be easily displaced or destroyed if it is not pinned by the notches, which we will discuss later in this paper.

We apply a single magnetic field pulse in the $+y$ direction to drive the dynamics of magnetic domains in the nanotrack. The magnetic field pulse strength is set to $B_{y}=100$ mT, and the pulse length is set to $\tau=200$ ps, which is applied during $t=20-220$ ps [see Fig.~\ref{FIG4}(b)].
It can be seen that the magnetization in the nanotrack rotate to the $+y$ direction driven by the magnetic field pulse during $t=20-220$ ps, and the magnetization continue to rotate mainly in the nanotrack plane after the pulse application and finally relax to a tail-to-tail magnetization configuration, as shown in Fig.~\ref{FIG4}(a) at $t=1000$ ps (see \blue{Supplemental Video 1} in Ref.~\onlinecite{SM}).
Such a tail-to-tail magnetization configuration is thus defined as the state ``1'' as it could be used to carry the binary information bit ``1''.
Therefore, the field-induced reorientation of magnetic domains leads to a smooth switching from state ``0'' to state ``1'', mimicking a unitary NOT operation that is fundamental to bit manipulation in spintronic information storage and processing devices.

The reorientation of the magnetic domains in the left half and right half of the nanotrack could be clearly indicated by the reduced in-plane magnetization components for the left half ($m_{x}^{\text{L}}$) and right half ($m_{x}^{\text{R}}$) of the nanotrack, respectively [see Fig.~\ref{FIG4}(c)].
During the field-driven reorientation and relaxation, both the in-plane magnetization component $m_y$ [see Fig.~\ref{FIG4}(d)] and out-of-plane magnetization component $m_z$ [see Fig.~\ref{FIG4}(e)] also vary with time, however, the variation amplitude of $m_z$ is much smaller as the magnetization lie in the $x$-$y$ plane.
The time-dependent total energy [see Fig.~\ref{FIG4}(f)], exchange energy [see Fig.~\ref{FIG4}(g)], demagnetization energy [see Fig.~\ref{FIG4}(h)], and Zeeman energy [see Fig.~\ref{FIG4}(i)] are given in Fig.~\ref{FIG4}. The total energy, exchange energy, and Zeeman energy decrease during the field-driven reorientation, while the demagnetization energy increases during the reorientation period. When the magnetic field pulse is turned off at $t=220$ ps, the total energy shows a sharp increase and then decreases to its initial value. The demagnetization energy also decreases to its initial value. Therefore, it can be seen that the formation of the tail-to-tail magnetization configuration after the pulse application is favored by the demagnetization effect, although the exchange energy will slightly increase due to the formation of a domain wall structure between the two square notches.

We also apply same in-plane magnetic field pulse to drive the system with the state ``1'' being the initial state, namely, the initial state is a tail-to-tail magnetization configuration identical to the one obtained at $t=1000$ ps in Fig.~\ref{FIG4}(a).
The geometric and material parameters are the same as that used in Fig.~\ref{FIG4} and the magnetic field pulse with $B_{y}=100$ mT and $\tau=200$ ps is also applied at the $+y$ direction.
The results are given in Fig.~\ref{FIG5}, where we find that the magnetic domain reorientation induced by the magnetic field pulse could also lead to the smooth switching from state ``1'' to state ``0''.
This means that the switching between state ``0'' and state ``1'' is reversible due to the magnetic domain reorientation induced by magnetic field pulses applied at the $+y$ direction.

Hence, as shown in Fig.~\ref{FIG6}, we demonstrate the reversible switching between state ``0'' and state ``1'' by applying a sequence of magnetic field pulses to drive the nanotrack with an initial state of head-to-head magnetization configuration (i.e., state ``0''), where the direction of magnetic field pulses is fixed at the $+y$ direction (see \blue{Supplemental Video 2} in Ref.~\onlinecite{SM}).
In Fig.~\ref{FIG6}, it can be seen that the reversible switching between state ``0'' and state ``1'' can be reliably achieved provided that the spacing between two adjacent pulses is long enough for the system to fully relax into a metastable state.

\begin{figure}[t]
\centerline{\includegraphics[width=0.49\textwidth]{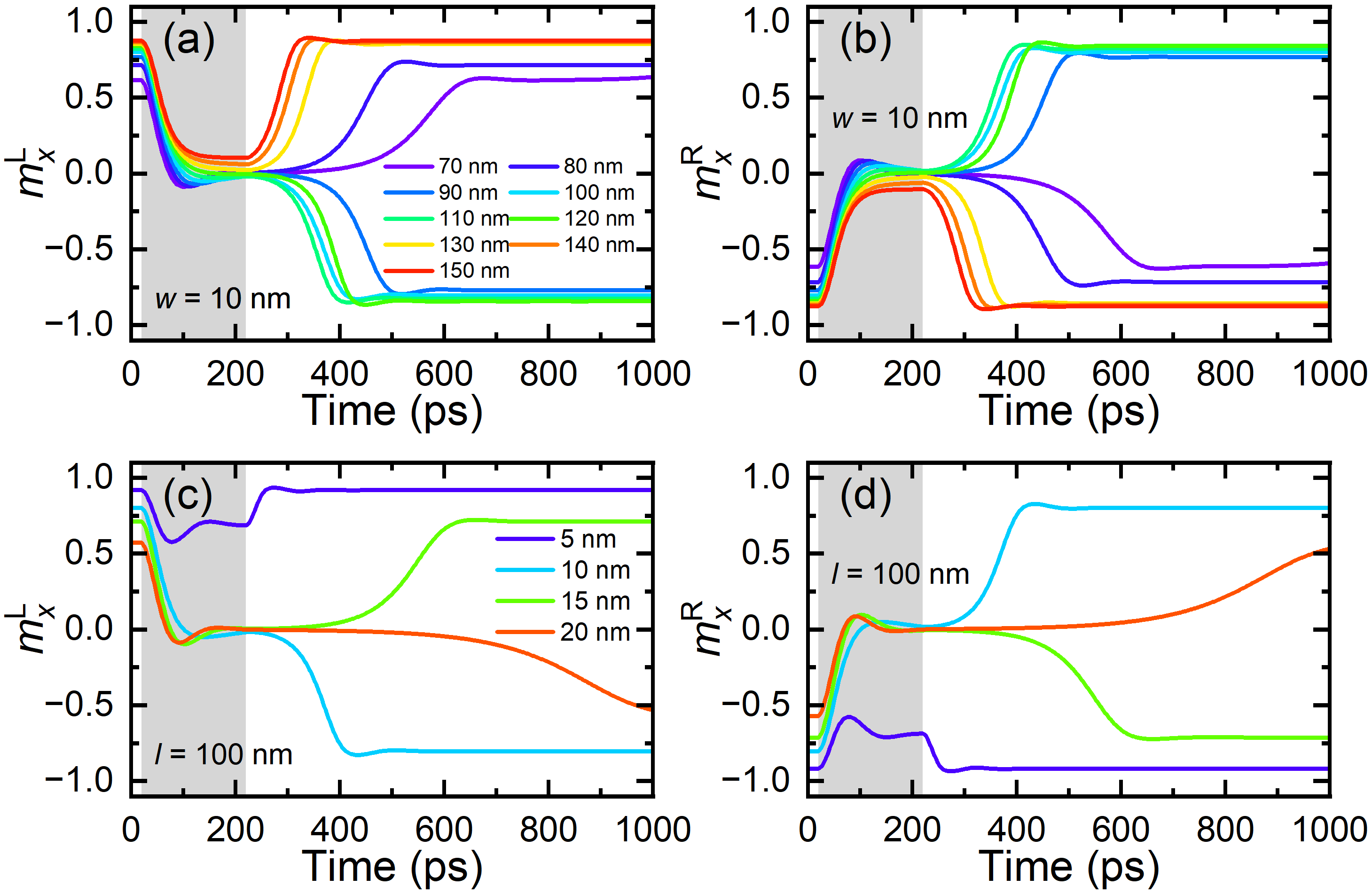}}
\caption{%
Dynamics of magnetic domains driven by a single magnetic field pulse in notched nanotracks with different geometric parameters.
(a) Time-dependent reduced in-plane magnetization component for the left half of the nanotrack $m_{x}^{\text{L}}$ for different track length $l$. The track width is fixed at $w=10$ nm.
(b) Time-dependent reduced in-plane magnetization component for the right half of the nanotrack $m_{x}^{\text{R}}$ for different $l$.
(c) Time-dependent $m_{x}^{\text{L}}$ for different track width $w$. The track length is fixed at $l=100$ nm.
(d) Time-dependent $m_{x}^{\text{R}}$ for different $w$.
Here, the initial relaxed state before the pulse application is the state ``0'' with a head-to-head magnetization configuration. The magnetic field pulse is applied at the $+y$ direction. The field strength $B_y=100$ mT, and the pulse length $\tau=200$ ps (i.e., applied during $t=20-220$ ps as indicated by the gray background). The system is relaxed until $t=1000$ ps.
}
\label{FIG9}
\end{figure}

\begin{figure*}[t]
\centerline{\includegraphics[width=0.99\textwidth]{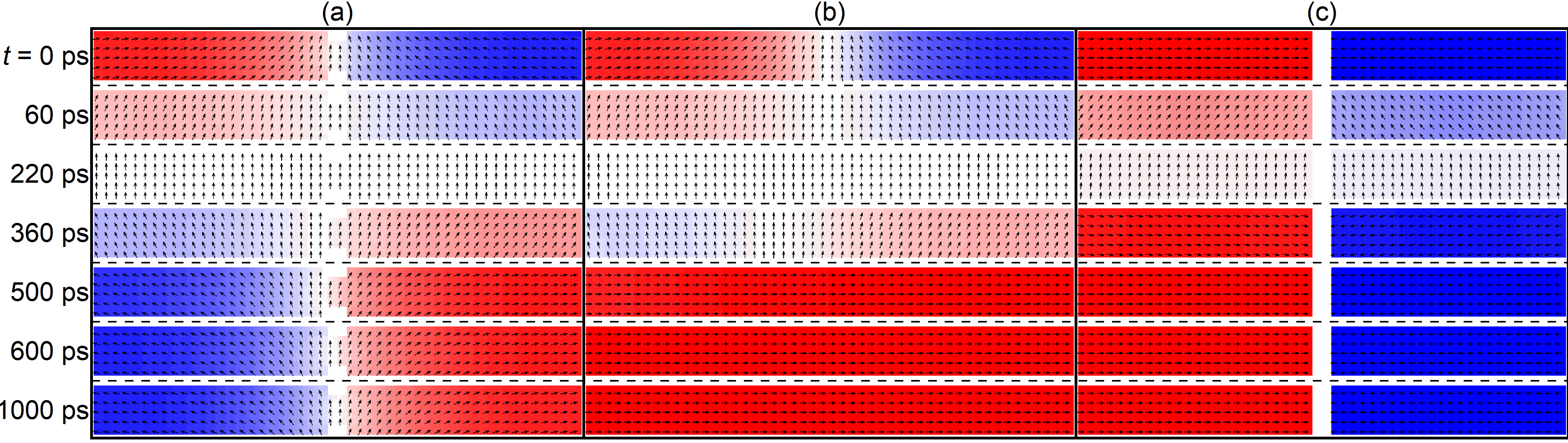}}
\caption{%
Field-induced reorientation of magnetic domains in nanotracks with and without the notches.
(a) Top-view snapshots showing the magnetization configurations in a default notched nanotrack at selected times. The arrows represent the magnetization directions. The in-plane magnetization component ($m_x$) is color coded: red is pointing at the $+x$ direction, blue is pointing at the $-x$ direction, and white is pointing at the $\pm y$ directions. The initial state ``0'' is a metastable head-to-head magnetization configuration, while the final state ``1'' is a metastable tail-to-tail magnetization configuration. The switching from state ``0'' to state ``1'' is realized by the magnetization reorientation driven by a magnetic field pulse applied at the $+y$ direction with a tiny component in the $+x$ direction. The magnetic field $\boldsymbol{B}=(0.005, 100, 0)$ mT, and the pulse length $\tau=200$ ps (i.e., applied during $t=20-220$ ps). The system is relaxed until $t=1000$ ps. Here, $\alpha=0.3$, $A=13$ pJ m$^{-1}$, and $M_\text{S}=860$ kA m$^{-1}$. The reference frame is the same as that in Fig.~\ref{FIG4}.
(b) Top-view snapshots showing the magnetization configurations in a nanotrack without notches at selected times. The length and width of the nanotrack are the same as that in (a). All material and magnetic field parameters are the same as that in (a). The initial state is a metastable head-to-head magnetization configuration, while the final state is a stable uniform magnetization configuration.
(c) Top-view snapshots showing the magnetization configurations in two nanotracks without notches at selected times. The system is the same as that in (a) but the area between the two notches is totally etched away. All material and magnetic field parameters are the same as that in (a). Treating the two nanotracks as a system, the initial state can be seen as a stable head-to-head magnetization configuration, while the final state is identical to the initial state.
}
\label{FIG10}
\end{figure*}

\subsection{Geometry and parameter dependence}
\label{se:Dependence}

In Fig.~\ref{FIG7}, we further study the field-induced reorientation dynamics of magnetic domains for different strengths and lengths of the applied magnetic field pulse.
The nanotrack geometry and magnetic parameters are the same as that used in Fig.~\ref{FIG4}. The initial relaxed state is a head-to-head magnetization configuration referred as the state ``0''. A single magnetic field pulse with strength $B_y$ and length $\tau$ is applied in the $+y$ direction and the system is relaxed after the pulse application until $t=1000$ ps.
It shows that the switching from state ``0'' to state ``1'' could be realized only when the field strength $B_y$ is larger than certain threshold value. For the magnetic field pulse with a suitable strength that is enough to induce the magnetic domain reorientation, only certain ranges of pulse lengths $\tau$ could result in the switching from state ``0'' to state ``1''. When a larger $B_y$ is applied, both the maximum and minimum values of $\tau$ required for the switching will be reduced as a general trend.

In Fig.~\ref{FIG8}, we explore the field-induced reorientation dynamics of magnetic domains for different intrinsic magnetic material parameters.
The nanotrack geometry and default magnetic parameters are the same as that used in Fig.~\ref{FIG4}. The initial relaxed state is a head-to-head magnetization configuration referred as the state ``0''. A single magnetic field pulse with strength $B_y=100$ mT and length $\tau=200$ ps is applied in the $+y$ direction during $t=20-220$ ps. The system is relaxed after the pulse application until $t=1000$ ps.
It is found that the damping parameter $\alpha$ may obviously affect the magnetic domain reorientation dynamics [see Figs.~\ref{FIG8}(a) and~\ref{FIG8}(b)], where we focus on the time-dependent reduced in-plane magnetization components for the left half $m_{x}^{\text{L}}$ and right half $m_{x}^{\text{R}}$ of the nanotrack. It shows that the field-induced switching from state ``0'' to state ``1'' could be realized when $\alpha=0.25-0.35$ or when $\alpha=0.05$. For relatively larger $\alpha=0.5$, the state ``0'' is not switched after the pulse application, although the magnetic domains are reorientated to the $+y$ direction by the field pulse around $t=220$ ps.

As shown in Figs.~\ref{FIG8}(c) and~\ref{FIG8}(d), the field-induced orientation of magnetic domains could lead to the a successful switching from state ``0'' to state ``1'' for a wide range of exchange constant. Similarly, the switching from state ``0'' to state ``1'' could also be realized for a wide range of saturation magnetization. These results suggest that the sensitivity of the field-induced reorientation of magnetic domains in the given nanotrack to damping parameter is higher than that to magnetic parameters, indicating the importance of the magnetization damping precession on the outcome of switching.

In Fig.~\ref{FIG9}, we also investigate the effects of nanotrack geometry on the field-induced reorientation dynamics of magnetic domains as well as the switching between state ``0'' and state ``1''.
The magnetic parameters are the same as that used in Fig.~\ref{FIG4}. The default nanotrack length is set as $l=100$ nm, and the default nanotrack width is set as $w=10$ nm.
We note that in this work, for the sake of simplicity, the length and width of the edge notch is fixed at $4$ nm and $2$ nm, respectively.
The initial relaxed state is a head-to-head magnetization configuration referred as the state ``0''. A single magnetic field pulse with strength $B_y=100$ mT and length $\tau=200$ ps is applied in the $+y$ direction during $t=20-220$ ps. The system is relaxed after the pulse application until $t=1000$ ps.
It can be seen from Figs.~\ref{FIG9}(a) and~\ref{FIG9}(b) that the field-induced reorientation of magnetic domains and its consequential switching from state ``0'' to state ``1'' are only realized in nanotracks with $l=90-120$ nm and $w=10$ nm.
Also, by fixing the nanotrack length at $l=100$ nm, the field-induced reorientation of magnetic domains and its consequential switching from state ``0'' to state ``1'' are only realized for nanotracks with $w=10$ nm or $w=20$ nm [see Figs.~\ref{FIG9}(c) and~\ref{FIG9}(d)].
The different shape of the nanotrack will result in different shape anisotropy due to the demagnetization effect, which may favor different magnetization configurations, and may also affect the field-induced reorientation and switching.

\begin{figure}[t]
\centerline{\includegraphics[width=0.49\textwidth]{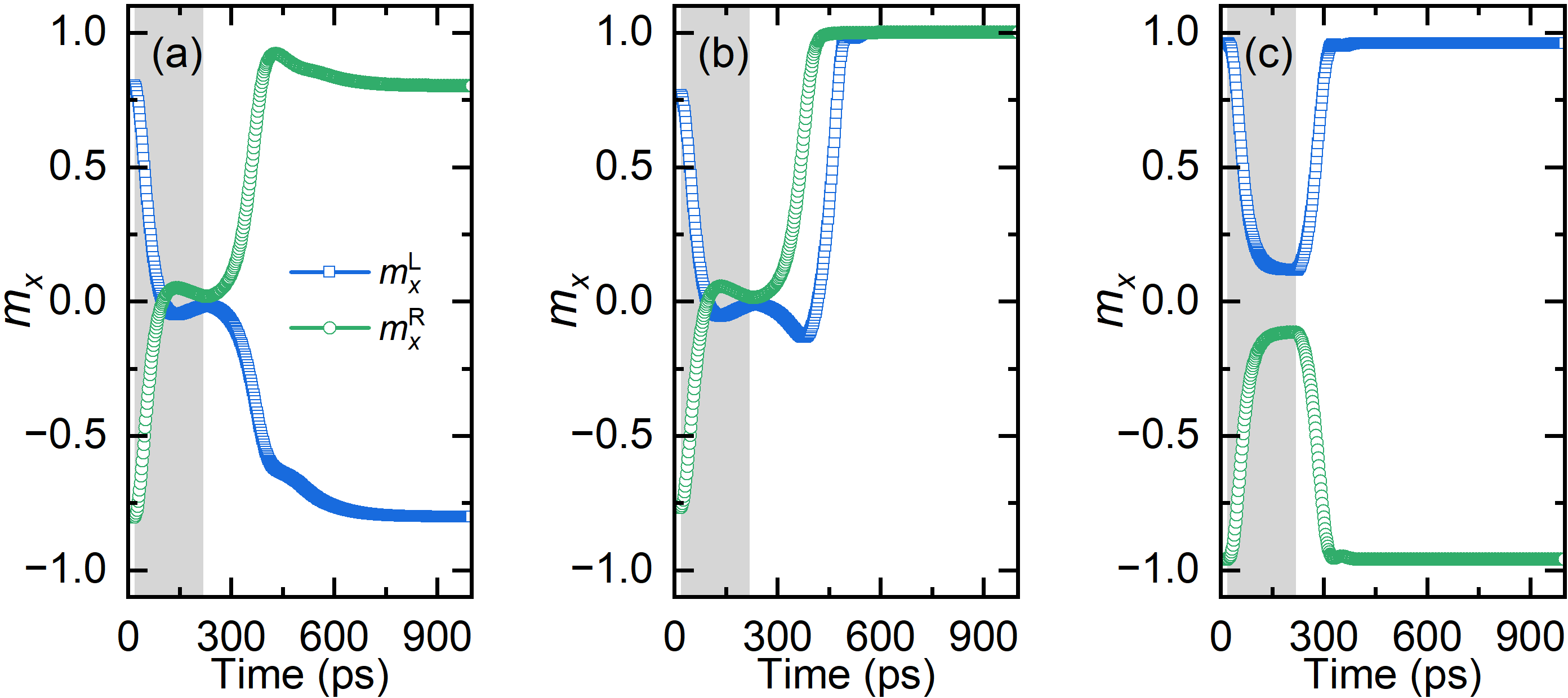}}
\caption{%
The reduced in-plane magnetization components for the left half ($m_{x}^{\text{L}}$) and right half ($m_{x}^{\text{R}}$) of the nanotrack system as functions of time.
(a) Time-dependent $m_{x}^{\text{L}}$ and $m_{x}^{\text{R}}$ for the system given in Fig.~\ref{FIG10}(a).
(b) Time-dependent $m_{x}^{\text{L}}$ and $m_{x}^{\text{R}}$ for the system given in Fig.~\ref{FIG10}(b).
(x) Time-dependent $m_{x}^{\text{L}}$ and $m_{x}^{\text{R}}$ for the system given in Fig.~\ref{FIG10}(c). The magnetic field pulse duration is indicated by the gray background.
}
\label{FIG11}
\end{figure}

\begin{figure}[t]
\centerline{\includegraphics[width=0.49\textwidth]{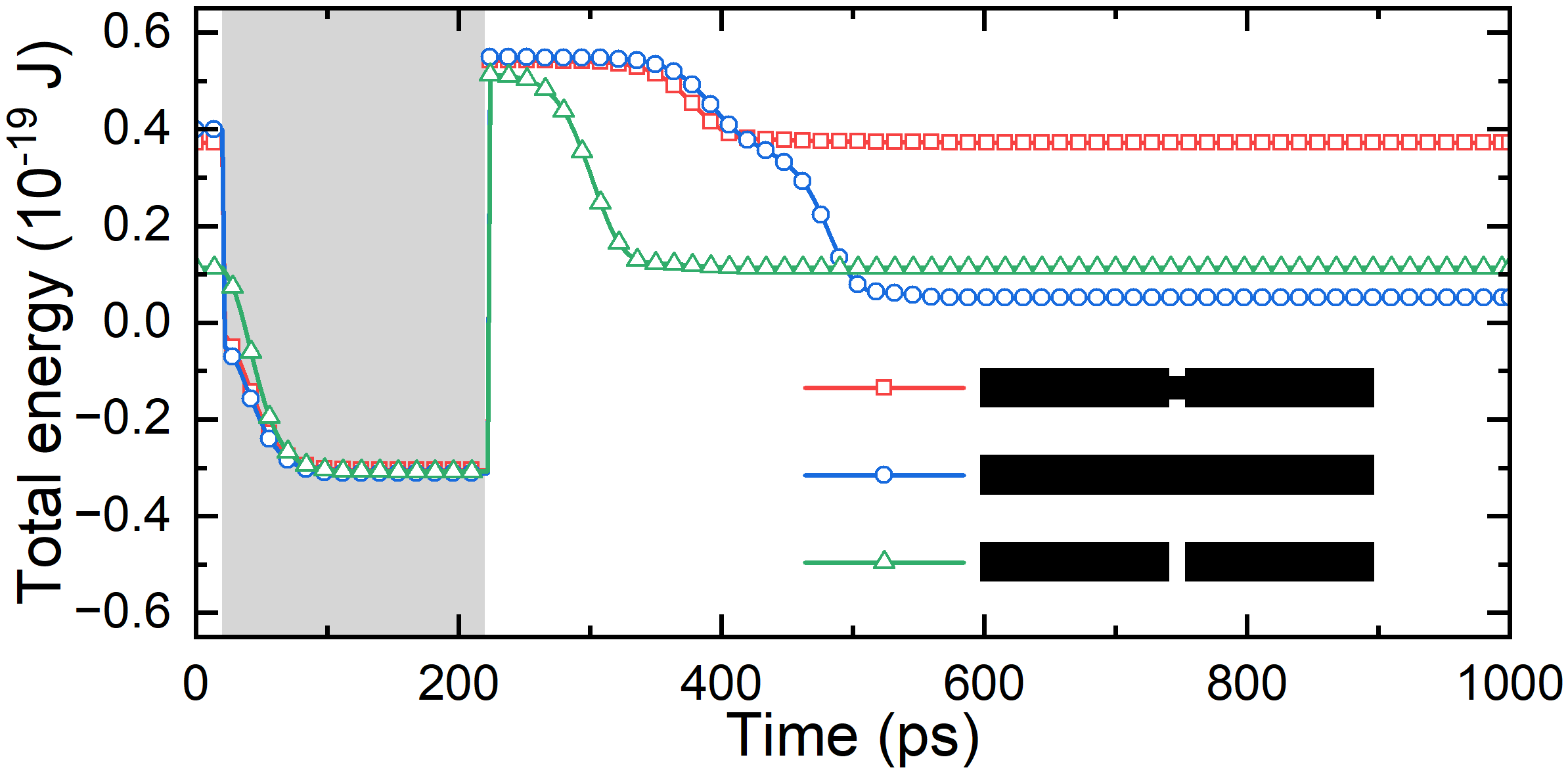}}
\caption{%
The total energies as a function of time for the three systems given in Fig.~\ref{FIG10}. The magnetic field pulse duration is indicated by the gray background.
}
\label{FIG12}
\end{figure}

\begin{figure}[t]
\centerline{\includegraphics[width=0.49\textwidth]{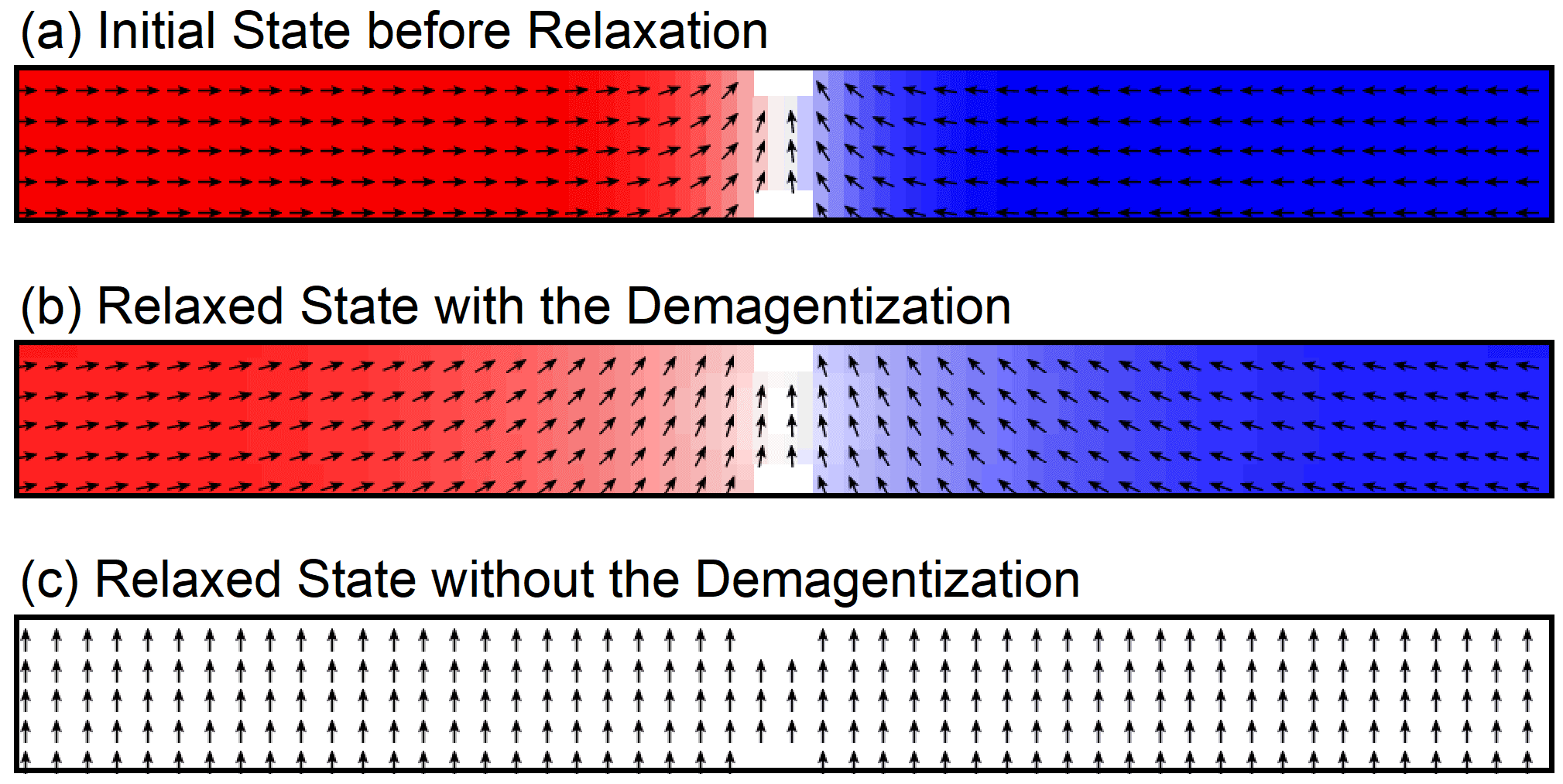}}
\caption{%
Top-view snapshots showing the initially relaxed magnetization configurations in a default notched nanotrack with and without the demagnetization effect.
(a) An ideal head-to-head magnetization configuration given as the initial state before the relaxation. The arrows represent the magnetization directions. The in-plane magnetization component ($m_x$) is color coded: red is pointing at the $+x$ direction, blue is pointing at the $-x$ direction, and white is pointing at the $\pm y$ directions.
(b) The relaxed state for the system with the demagnetization effect is a metastable head-to-head magnetization configuration.
(c) The relaxed state for the system without the demagnetization effect is a uniform state magnetized along the $+y$ direction.
}
\label{FIG13}
\end{figure}

\subsection{Effect of the notches}
\label{se:Notches}

For the purpose of elucidating the effect of the notches on the field-induced magnetic domain reorientation dynamics in the nanotrack, we carry out three simulations with different nanotrack geometries to compare with each other.
In the first simulation, as shown in Fig.~\ref{FIG10}(a), the nanotrack geometry and magnetic parameters are the same as that used in Fig.~\ref{FIG4}. The initial relaxed state is a head-to-head magnetization configuration referred as the state ``0''.
A single magnetic field pulse of $\boldsymbol{B}=(0.005, 100, 0)$ mT is applied to drive the magnetization dynamics in the nanotrack. Namely, the magnetic field is applied mainly at the $+y$ direction but with a tiny component in the $+x$ direction. The magnetic field is applied for $200$ ps during $t=20-220$ ps, and then the system is relaxed after the pulse application until $t=1000$ ps.
We note that we consider a tiny component of magnetic field in the $x$ direction to mimic an additional tiny perturbation in the field pulse, which would highlight the effect of the notches in enhancing the stability of the domain wall configuration during the reorientation.
It can be seen that the applied magnetic field pulse drives the reorientation of magnetic domains, and leads to the switching from state ``0'' to state ``1'' after the relaxation (see \blue{Supplemental Video 3} in Ref.~\onlinecite{SM}).
The switching from state ``0'' to state ``1'' can also be seen from the time-dependent $m_{x}^{\text{L}}$ and $m_{x}^{\text{R}}$ of the system [see Fig.~\ref{FIG11}(a)].

However, the snapshots selected at different times in Fig.~\ref{FIG10}(a) show that during the magnetic domain reorientation the domain wall structure at the center of the nanotrack first moves slightly toward the $-x$ direction (i.e., away from the notches), and then moves toward the $+x$ direction (i.e., back to the notches) [cf. Figs.~\ref{FIG4}(a) and~\ref{FIG10}(a)].
The domain wall motion toward the $-x$ direction is caused by the tiny $+x$ component of the applied magnetic field pulse, while its motion toward the $+x$ direction is due to the attractive interaction between the domain wall and the notches.

We note that such a displacement of the domain wall from the center of the nanotrack could result in the instability of the head-to-head and tail-to-tail magnetization configurations in a clean nanotrack without any pinning effect.
For example, as shown in Fig.~\ref{FIG10}(b), if we remove the notches from the center of the nanotrack while keeping other parameters unchanged, the magnetic field pulse will result in the transition from the initial head-to-head magnetization configuration to a uniform ferromagnetic state by driving the domain wall structure out of the nanotrack (see \blue{Supplemental Video 4} in Ref.~\onlinecite{SM}).
The transition is also indicated by the time-dependent $m_{x}^{\text{L}}$ and $m_{x}^{\text{R}}$ of the system [see Fig.~\ref{FIG11}(b)].
Here we point out that the reorientation could, in principle, be realized in a nanotrack without the notches under the exact same protocol used in Fig.~\ref{FIG4}, i.e., $\boldsymbol{B}=(0, 100, 0)$ mT. However, this is only true when the head-to-head or tail-to-tail domain wall is initially placed at the exact center of the nanotrack, and the applied magnetic field is perfectly aligned perpendicularly to the nanotrack. Otherwise, the head-to-head or tail-to-tail domain wall could easily be destroyed as demonstrated in Fig.~\ref{FIG10}(b).

We also apply a same magnetic field pulse of $\boldsymbol{B}=(0.005, 100, 0)$ mT to drive the system where the area between two notches is totally etched away, as shown in Fig.~\ref{FIG10}(c). It can be seen that the initial and relaxed final states of the system remain unchanged after the application of the magnetic field pulse (see \blue{Supplemental Video 5} in Ref.~\onlinecite{SM}). The magnetization dynamics in the left and right parts of the system are also described by the time-dependent $m_{x}^{\text{L}}$ and $m_{x}^{\text{R}}$, respectively [see Fig.~\ref{FIG11}(c)].

In Fig.~\ref{FIG12}, we show the time-dependent total energies for the three simulations with different nanotrack geometries discussed above.
It shows that the nanotrack with two notches at the upper and lower edges of the nanotrack center can realize the field-induced switching between two metastable states with the same energy (i.e., the head-to-head and tail-to-tail magnetization configurations). In the nanotrack without notches, the final state is ground state with an energy lower than that of the initial metastable state. In the nanotrack system with the area between two notches being totally etched away (i.e., a system of two shorter nanotracks), the initial and final states are ground states with the energy close to the ground state in the longer nanotrack without notches.
We also point out that the demagnetization effect plays an important role in the stabilization of the head-to-head or tail-to-tail magnetization configuration in our model. As shown in Fig.~\ref{FIG13}, if we turn off the demagnetization in the simulation, the initial head-to-head magnetization configuration cannot be relaxed and will evolve into an uniform state during the initial relaxation before the application of the magnetic field pulse.
The results discussed in this section suggest that the notches at the nanotrack center can pin the domain wall structure, and thus, enhance the stability of the head-to-head and tail-to-tail magnetization configurations, which are important for the realization of the fixed field-induced switching between state ``0'' and state ``1''.

\subsection{Implications for classical and quantum applications}
\label{se:Implications}

The reversible switching between state ``0'' and state ``1'' in the nanotrack without changing the driving field direction may have important implications for both classical and quantum spintronic applications.
For example, such a controlled reversible single-bit flipping operation driven by magnetic field applied in a fixed in-plane direction could be used to build a memory device, where information is stored in arrays of nanotracks with one bit per track, similar to the bit-patterned magnetic recording architecture~\cite{Albrecht_2015}.

On the other hand, recent studies have suggest that nanoscale magnetic textures, including domain walls~\cite{Zou_2022}, merons~\cite{Xia_CM2022}, and skyrmions~\cite{Psaroudaki_PRL2021,Psaroudaki_PRB2022,Xia_PRL2023}, can be used as building blocks in quantum computation.
As the reversible switching between state ``0'' and state ``1'' in the given system could be realized without reversing the sign of the applied in-plane magnetic field, the system may also, in principle, make it possible to build a quantum Pauli-X gate as long as the size of the nanotrack is of a few nanometers~\cite{Zou_2022} in real experiments.
Namely, magnetic textures with dimensions down to a few nanometers could be quantum objects~\cite{Lohani_PRX2019,Psaroudaki_PRL2021,Psaroudaki_PRB2022,Siegl_PRR2022,Zou_2022,Xia_CM2022,Xia_PRL2023}.

Note that the Pauli-X gate is a quantum version of the classical NOT gate, which transforms the pure state $\left|0\right\rangle$ to $\left|1\right\rangle$ and vice versa.
Namely, the Pauli-X gate is an operation that can be expressed by the Pauli matrix $\sigma_{x}$ acting on a single qubit, for example, $\sigma_{x}|0\rangle=|0\rangle\langle 1|0\rangle+|1\rangle\langle 0|0\rangle=|1\rangle$ and $\sigma_{x}|1\rangle=|0\rangle\langle 1|1\rangle+|1\rangle\langle 0|1\rangle=|0\rangle$, where $\sigma_{x}$ may correspond to an in-plane magnetic field with a fixed direction in real experiments.
It should be noted that it is not required to fix the direction of the driving field in the operation of the classical NOT gate.

Besides, for a classical system, the reversible switching between state ``0'' and state ``1'' induced by magnetic fields with a fixed direction may simplify the bit flipping operation as only the strength of the magnetic field needs to be programmed.
We note that the magnetic field applied in the device plane could be realized by using a micro coil covering the top and bottom surfaces of the nanotrack. One may also fabricate a nanotrack parallel and attached (i.e., underneath or beyond) to a flat current wire, so that the Oersted field in the in-plane direction could be used to drive the magnetization dynamics in the nanotrack.

\begin{figure}[t]
\centerline{\includegraphics[width=0.49\textwidth]{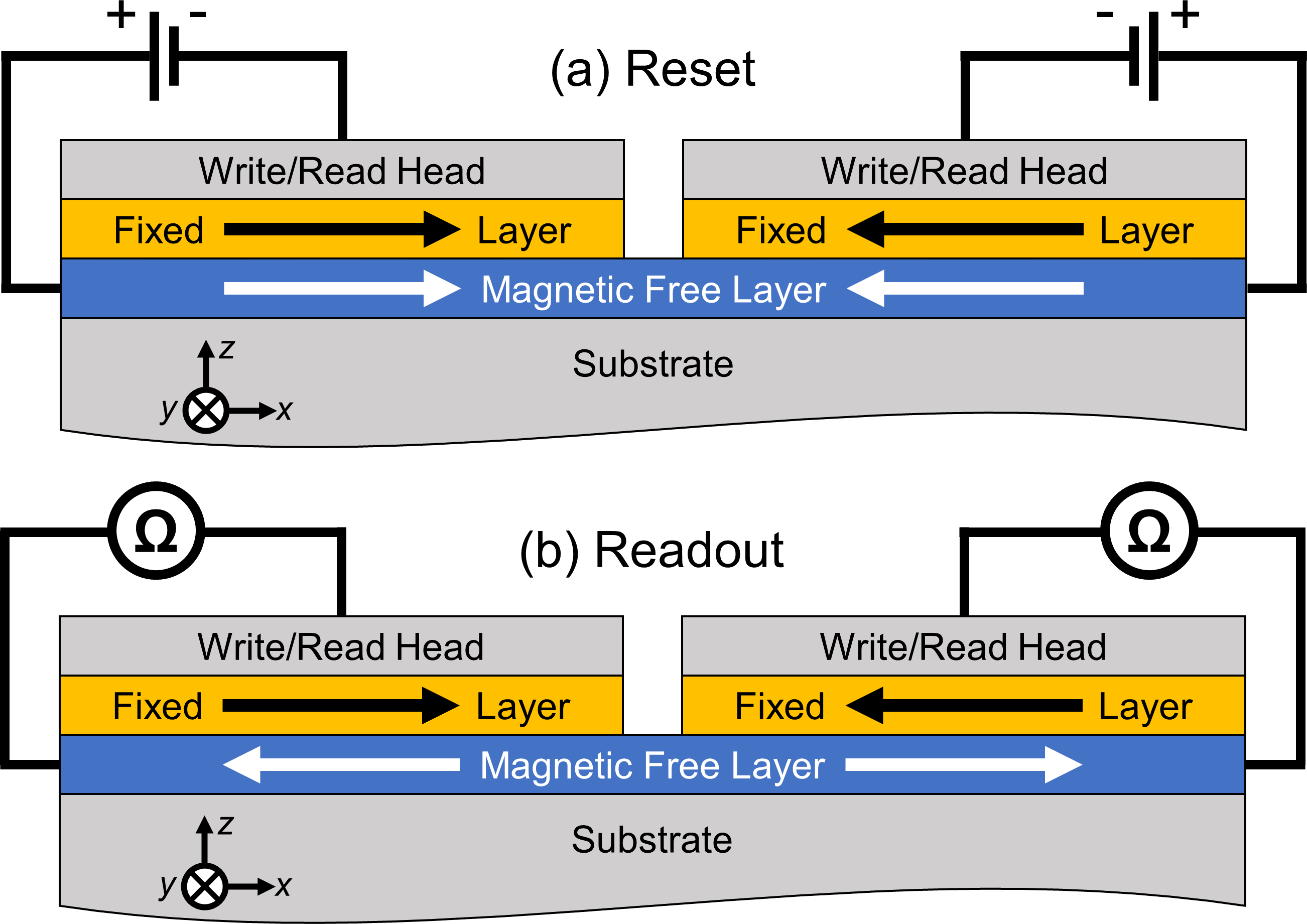}}
\caption{%
Schematic illustrations showing possible experimental setups for the initialization, readout, and reset operations.
(a) The initialization and reset functions could be realized by using two magnetic tunnel junction (MTJ) elements with fixed layers placed upon the left and right parts of the notched ferromagnetic nanotrack (i.e., free layers), respectively.
(b) The readout operation could be realized by using the above-mentioned two MTJ elements as the readout sensors.
}
\label{FIG14}
\end{figure}

In Fig.~\ref{FIG14}, we also provide two schematic illustrations showing possible experimental setups for the initialization, readout, and reset operations. First, the initial head-to-head magnetization configuration in the nanotrack could be created by using two magnetic tunnel junction (MTJ) elements placed upon the left and right parts of the notched nanotrack, where the fixed layers in the two MTJ elements have opposite uniform magnetization configurations as depicted in Fig.~\ref{FIG14}. By injecting vertical spin currents through the two MTJ elements, one should be able to create or reset the head-to-head magnetization configuration in the notched nanotrack. Besides, the two MTJ elements could also be used as the readout sensors to detect the tail-to-tail magnetization configuration in the notched nanotrack after the magnetic domain reorientation.

\section{Conclusion}
\label{se:Conclusion}

In conclusion, we have studied the reorientation of head-to-head and tail-to-tail magnetic domains in a notched nanotrack driven by in-plane magnetic field pulses, where the direction of magnetic field is fixed at the $+y$ direction, i.e., perpendicular to the length direction of the nanotrack.
It is found that the field-induced reorientation of magnetic domains could lead to reversible switching between a head-to-head magnetization configuration and a tail-to-tail magnetization configuration.
The head-to-head and tail-to-tail magnetization configurations in the nanotrack could be treated as binary bit states ``0'' and ``1'', respectively. Thus, the reversible switching between state ``0'' and state ``1'' could be induced by applying in-plane magnetic field pulses perpendicular to the nanotrack without changing the direction of the magnetic field.
Such a phenomenon is in stark contrast to typical field-driven and current-driven domain switching and domain wall motion, where the reversible switching and motion are obtained by reversing the sign of driving force.
The reorientation dynamics of magnetic domains depends on both the intrinsic magnetic parameters, the nanotrack geometry, and the applied magnetic field pulse profile. The damping parameter and nanotrack geometry may play important roles on the reorientation dynamics and the consequential switching between state ``0'' and state ``1'', however, the overall dynamic process (i.e., including the field-driven reorientation and zero-field relaxation) is robust to the variations of magnetic parameters, including the exchange constant and saturation magnetization.
Namely, the reorientation operation is robust for a wide parameter range of exchange constant and saturation magnetization, although the suitable parameter spaces for the reorientation are not extremely wide for parameters such as the damping parameter.
It should be emphasized that the notches at the center of the upper and lower nanotrack edges could enhance the stability and reliability of the head-to-head and tail-to-tail magnetization configurations, especially in the presence of a perturbation in the driving field.
Our results could be important for understanding the field-driven dynamics of magnetic domains in narrow nanotracks with notches, and may also provide a new way for the design of novel spintronic devices.
We also believe that our computational findings will stimulate more theoretical and experimental efforts to explore novel magnetization dynamics in complex and artificial nanostructures.


\begin{acknowledgments}
X.Z. and M.M. acknowledge support by CREST, the Japan Science and Technology Agency (Grant No. JPMJCR20T1).
M.M. also acknowledges support by the Grants-in-Aid for Scientific Research from JSPS KAKENHI (Grant No. JP20H00337).
J.X. was a JSPS International Research Fellow supported by JSPS KAKENHI (Grant No. JP22F22061).
O.A.T. acknowledges support by the Australian Research Council (Grant No. DP200101027), the Russian Science Foundation (Grant No. 21-42-00035), the Cooperative Research Project Program at the Research Institute of Electrical Communication, Tohoku University (Japan), and by the NCMAS grant.
G.Z. acknowledges support by the National Natural Science Foundation of China (Grants No. 51771127, No. 51571126, and No. 51772004), and Central Government Funds of Guiding Local Scientific and Technological Development for Sichuan Province (Grant No. 2021ZYD0025).
Y.Z. acknowledges support by the Guangdong Basic and Applied Basic Research Foundation (Grant No. 2021B1515120047), the Guangdong Special Support Project (Grant No. 2019BT02X030), the Shenzhen Fundamental Research Fund (Grant No. JCYJ20210324120213037), the Shenzhen Peacock Group Plan (Grant No. KQTD20180413181702403), the Pearl River Recruitment Program of Talents (Grant No. 2017GC010293), and the National Natural Science Foundation of China (Grant No. 11974298).
X.L. acknowledges support by the Grants-in-Aid for Scientific Research from JSPS KAKENHI (Grants No. JP20F20363, No. JP21H01364, No. JP21K18872, and No. JP22F22061).
M.E. acknowledges support by the Grants-in-Aid for Scientific Research from JSPS KAKENHI (Grant No. JP23H00171).
M.E. also acknowledges support by CREST, JST (Grant No. JPMJCR20T2).
\end{acknowledgments}


M.E. and X.Z. conceived the idea. M.M. and X.L. coordinated the project. X.Z. and J.X. performed the computational simulation and theoretical analysis supervised by M.M., X.L., and M.E. The manuscript is drafted by X.Z. and J.X., and revised with input from O.A.T., G.Z., Y.Z., M.M., X.L., and M.E. All authors discussed the results and reviewed the manuscript. X.Z. and J.X. contributed equally to this work.



\end{document}